# Dipole-dipole correlations in the nematic phases of symmetric cyanobiphenyl dimers and their binary mixtures with 5CB

Evangelia E. Zavvou, *[a] Efthymia Ramou,[a] Ziauddin Ahmed,[b] Chris Welch,[b] Georg H. Mehl,[b] Alexandros G. Vanakaras[c] and Panagiota K. Karahaliou[a]

[a] *Department of Physics, University of Patras, 26504 Patras, Greece.*
[b] *Department of Chemistry, University of Hull, HU6 7RX, UK.*
[c] *Department of Materials Science, University of Patras, 26504 Patras, Greece.*

## Abstract

We report on the temperature dependence of birefringence and of the static dielectric permittivity tensor in a series of binary mixtures between the symmetric, bent-shaped, 1′′,9′′-bis(4-cyanobiphenyl-4′-yl)nonane (CB9CB) dimer and the monomeric nematogen 5CB. In the studied composition range the mixtures exhibit two nematic phases with distinct birefringence and dielectric features. Birefringence measurements are used to estimate the temperature dependence of the tilt between the axis defining the nanoscale helical modulation of the low temperature nematic phase with the (local) direction of the maximal alignment of the cyanobiphenyl units. Planar as well as magnetically and/or electrically aligned samples are used to measure the perpendicular and parallel components of the dielectric permittivity in both nematic phases. A self-consistent molecular field theory that takes into account flexibility and symmetry of the constituent mesogens is introduced for the calculation of order parameters and *intra*-molecular orientational dipolar correlations of the flexible dimers as a function of temperature/concentration. Utilising the tilt angle, as calculated from the birefringence measurements, and the predictions of the molecular theory, dielectric permittivity is modelled in the framework of the anisotropic version of the Kirkwood-Fröhlich theory. Using the *inter*-molecular Kirkwood correlation factors as adjustable parameters, excellent agreement between theory and permittivity measurements across the whole temperature range and composition of the mixtures is obtained. The importance of the orientational, *intra*- and *inter*-molecular, dipolar correlations, their relative impact on the static dielectric properties, as well as their connection with the local structure of the nematic phases of bent-shaped bimesogens, is discussed.

## Introduction

Uniaxial nematics represent the simplest kind of mesomorphic self-organisation in a fluid system comprised molecules with shape anisometry.[1] For more than a century, the only known nematic phases were the conventional nematic phase (N), formed by achiral mesogens, while in the presence of molecular chirality blue phases (BP) and/or the cholesteric nematic phase (N*) emerge. Over the span of the last twenty years, the notion of the nematic polymorphism in thermotropic liquid crystals has dramatically expanded due to the discovery of new fascinating states of matter with nematic ordering. These include the biaxial nematic phase,[2] the periodically modulated nematic phase of achiral bent-core dimers[3] and the recently discovered ferroelectric nematic phase.[4]

The identification of a first order nematic-nematic phase transition dates back to 2010, when a careful reexamination[3] of the phase behaviour of methylene-linked liquid crystal dimers with odd number of carbon atoms in the spacer, led to one of the most fascinating discoveries in the LC science: the observation of spontaneous structural chirality in a nematic phase formed by achiral molecules.[5,6] The chirality of the novel low temperature nematic phase, initially termed as $N_x$, was confirmed by NMR studies[7–9] and, more recently, by Circular Dichroism spectroscopy.[10] The orientational order within the $N_x$ is periodically modulated with an extremely short pitch in the order of 10 nm, initially evaluated through freeze-fracture TEM/AFM[5,11,12] and directly measured though Resonant X-Ray Scattering experiments.[13–17]

Soon after its discovery, the $N_x$ phase was identified with the theoretically predicted twist-bend nematic phase ($N_{tb}$),[18] which was originally proposed by R. B. Meyer[19] as a possible spontaneous macroscopic deformation mode in locally polar nematics. Later, Dozov,[20] on the basis of nematic elasticity, demonstrated that a spontaneous twist-bend deformation may indeed be stable in systems of achiral bent-core molecules. Based, also, on nematic elasticity, several other theoretical models have been developed.[21–24] A different interpretation of the origins of the nanoscale modulation was proposed later, according to which the structure of the $N_x$ phase is connected with a genuine, entropically driven molecular ordering, which corresponds to a locally polar structural organisation. As a result, $N_x$ is characterised by a local polar director, that roto-translates generating a molecular length-scale 1-D modulation, corresponding to the so-called polar-twisted nematic, $N_{PT}$.[25–28] Yet, there is not a unified and broadly accepted interpretation of the microscopic origins and of the nature of the thermodynamic driving forces dictating the emergence of the $N_x$ phase, as will be designated in this work.

The archetypal molecular structures exhibiting the $N_x$ phase, are the symmetric CBnCB [3,18,29–33] dimers with odd number of carbon atoms in the flexible spacer. Even members of the homologous series exhibit only the conventional N phase, which represents another manifestation of the importance of the extensively studied odd-even effects in liquid crystals.[34–36] The dependence of the mesomorphic behaviour of such structures on the parity and the length of the flexible spacer is a consequence of the dominant molecular conformations, which in the case of the odd-membered dimers are bent and in the case of even-membered dimers linear. Thus, it is commonly accepted that an overall bent molecular shape is a necessary, although not always sufficient, condition for the formation of nanomodulated nematic phases.[37,38] Extensive research on this novel form of nematic organisation has led to the discovery of a rich variety of molecular architectures exhibiting the $N_x$ in addition to the conventional N phase. These include asymmetric dimers, oligomers, polymers, rigid bent-core mesogens and hydrogen-bonded supramolecular systems.[39]

In terms of symmetry, it is also known that, despite the rather distinct differences concerning the structure of the $N_x$ and N phases at the nanoscale, on larger scales both nematics appear uniaxial. Consequently, macroscopic 2$^{nd}$ rank tensor properties, as well as their corresponding anisotropies, are of special interest for the study of the structure-properties relationship of the two nematic phases. For example, birefringence in the high-temperature nematic phase of $N_x$-forming dimers increases with decreasing temperature, following a Haller-type[40] temperature dependence, similarly to low molar mass nematogens. On the contrary, on entering the low-temperature $N_x$ phase, $\Delta n$ decreases strongly on cooling, a behaviour associated with the nanoscopic helical modulation.[41,42] Concerning the dielectric anisotropy of odd-membered CBnCB dimers, an abrupt increase is observed at the isotropic-nematic phase transition, followed by a smoothly decreasing trend,[18,29,30,43,44] clearly opposed to the corresponding behaviour of $\Delta\varepsilon$ of cyanobiphenyl monomers.[45] Across the N-$N_x$ transition, only subtle variations of the $\varepsilon_\parallel$ and $\varepsilon_\perp$ components are observed, followed by a steeper, compared to the N phase, reduction of $\Delta\varepsilon$. This diminishment might eventually lead to sign reversal of $\Delta\varepsilon$ deeply in the $N_x$ phase.[18,30]

Dielectric studies in the high temperature N phase of binary systems between CBnCB dimers and cyanobiphenyl monomers report a systematic increase of $\Delta\varepsilon$ upon increasing the monomer content towards the behaviour of the monomeric system. Interestingly, the added presence of the monomer reveals distinct differences in the temperature dependence of both permittivity components with the onset of the N-$N_x$ transition, which are more pronounced in the $\varepsilon_\perp$ component.[46,47] Specifically, the smoothly decreasing trend of $\varepsilon_\perp$ of the neat dimer alters in mixtures, exhibiting a significant increase after the N-N$_x$ transition, the magnitude and the temperature range of which depend on the monomer concentration. These differences are certainly connected to the helix formation, however, it seems that the magnitude of the electric dipole moment of the added mesogen plays also a significant role, since similar trends have been reported in CB7CB/FFO9OCB binary mixtures, with FFO9OCB bearing a strong longitudinal dipole moment in one of its mesogenic cores.[48]

The dielectric anisotropy in the case of CBnCB series [18,30,43,44] has been associated with the temperature dependent conformational statistics of the dimer.[49,50] Characteristically, a jump of $\varepsilon_\parallel$ is observed at the IN transition in odd CBnCB dimers, which has been attributed to the stabilisation of hairpin conformers, which significantly contribute to the mean square dipole moment parallel to the director. Nevertheless, the increase of the orientational order deeper in the N phase favours the extended conformers with a statistically lower net longitudinal dipole moment, leading to monotonically decreasing values of $\varepsilon_\parallel$, on further cooling.

For a more in-depth analysis, the short-range *inter*-molecular dipolar correlations, usually overseen in the interpretation of the dielectric properties of the dimeric systems, should also be considered.[50] Actually, these correlations are known to significantly affect the dielectric behaviour of the corresponding monomeric nCB systems, where the nitrile (CN) groups tend to associate in an antiparallel fashion. Such type of dipolar association is already present within the isotropic phase of the nCB monomers, as demonstrated by a pretransitional decrease of $\varepsilon_{iso}$ close to $T_{NI}$,[51] while on entering the N phase the antiparallel associations are significantly enhanced.[52,53] Additionally, the non-typical dielectric behaviour observed in some systems of strongly polar rod-like[54] and bent-core molecules,[55] especially in the vicinity of N-SmA transition, has been interpreted in terms of *inter*-molecular dipole correlations.

In this work, we have performed systematic measurements of the static dielectric permittivity of CB9CB dimer and of CB9CB/5CB binary mixtures. Measurements span temperatures ranging from the isotropic down to temperatures deeply in the N$_x$ phase. Birefringence measurements have also been conducted in the same temperature range. We rationalize our experimental findings with the help of the well-established extension of Kirkwood-Fröhlich theory for dielectric permittivity of anisotropic polar fluids. To do this, we introduce a simple mean field model for flexible bent-core molecules, as well as for its mixtures with rigid monomers. With this model, we are able to calculate the temperature dependence of the orientational order parameters, as well as of the *intra*-molecular orientational dipolar correlations. With these ingredients, combining theory and permittivity measurements, we determine the *inter*-molecular Kirkwood correlation factors associated with the dipolar correlations of cyano-groups belonging to different molecules. These factors play a key role on the excellent representation of the dielectric permittivity measurements across the whole temperature and composition range.

The paper is organized as follows: in the next section we present in detail the experimental protocols implemented for the optical and dielectric characterisation. Next, we present and discuss our experimental findings. We continue with a detailed presentation of the theoretical model and the main assumptions involved. In the same section, we discuss thoroughly the role of the Kirkwood correlation factors on the success of the theory to reproduce the experimentally determined dielectric permittivities. The major conclusions are summarised in the last section.

**Experimental**

**Materials and sample preparation**

The studied liquid crystal dimer 1'',9''-bis(4-cyanobiphenyl-4'-yl) nonane (CB9CB) was synthesized in the Department of Chemistry of the University of Hull (UK). Binary mixtures were prepared between CB9CB and its corresponding monomer 4'-pentyl-4-biphenylcarbonitrile (5CB) (Sigma-Aldrich, Merck). The chemical structure and phase sequences of both compounds are presented in Fig. 1. Binary mixtures were prepared by co-dissolution of pre-weighed amounts of each compound in dichloromethane, followed by 1 min sonication. Then, the solutions were heated at 60°C for about 1 hour, allowing for complete solvent evaporation.

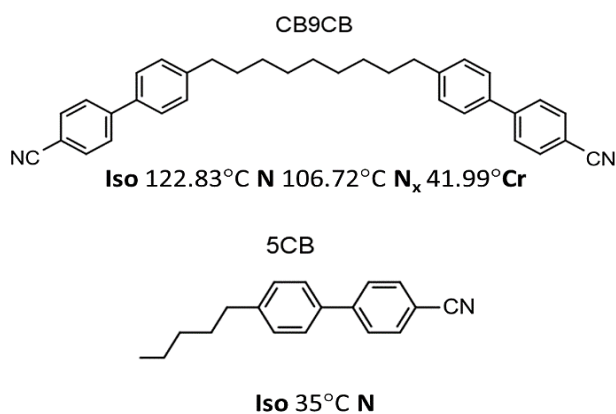

Fig. 1 Chemical structure and transition temperatures on cooling of CB9CB dimer31 and 5CB.

**POM observations and Birefringence**

Optical characterisation of pure compounds and mixtures was carried out on a Zeiss Axioskop 40 pol Polarizing Optical Microscope equipped with a Linkam LTS420 hotstage and a ProgRes CT5 camera. Optical observations in untreated glass slides upon cooling at a rate of 10°C/min were used to verify phase sequences and transition temperatures already present in the literature.[59] Optical retardation ($\Gamma$) measurements were carried out in planar cells with 5 μm spacing (LCC5 from Linkam, anti-parallel rubbing) using a Berek compensator (Leitz) and monochromatic light of $\lambda = 546$ nm. The thickness of empty cells ($d = 5.00 \pm 0.01$ μm) was determined by White Light Reflectance Spectroscopy using an FR-series tool from Theta-Metrisis (Greece). For the calculation of birefringence ($\Delta n$), the measured optical retardation ($\Gamma$) was divided by the corresponding cell thickness.

**Dielectric studies**

The dielectric response of pure compounds and binary mixtures was measured in the frequency range from 100 Hz to 1 MHz employing an Alpha-N Frequency Response Analyser (Novocontrol, Germany). The components of the complex dielectric permittivity, $\varepsilon^*_\perp(\omega)$ and $\varepsilon^*_\parallel(\omega)$, of 5CB were acquired in commercial planar (20 $\mu m$, Instec) and homeotropic (18.3 $\mu m$, Instec) cells, respectively, using 0.2 $V_{rms}$ probe field. For the dimer and mixtures, $\varepsilon^*_\perp(\omega)$ was measured in 20 $\mu$m planar cells (Instec, antiparallel rubbing), on cooling from the isotropic phase, using a 0.5 $V_{rms}$ probe field, which lies well below the threshold of Fréedericksz transition. The capacitance of the empty cells was determined prior to sample preparation. During measurements, the samples were held in a Novocontrol cryostat and temperature was controlled and stabilized within ±0.02°C by a Quatro Cryosystem temperature controller (Novocontrol).

For the determination of the parallel component $\varepsilon^*_\parallel(\omega)$ two different protocols were followed within the N and N$_x$ phase, respectively. In the high temperature conventional N phase, homeotropic alignment of the director was achieved using a magnetic field of $B = 1.4$T, on cooling from the isotropic phase (*Protocol 1 -P.1*). For this purpose, the 20 $\mu$m planar cells were placed in a homemade sample holder between the Helmholtz coils of an electromagnet. The magnetic field was applied along the cell normal and after a 50s waiting period, dielectric spectra were acquired with a 0.5 $V_{rms}$ sinusoidal field (Alpha-N analyser, Novocontrol). During the magnetic field experiments, temperature was controlled using an ITC502S Oxford Instruments temperature controller allowing for temperature stabilisation better than ±0.1°C. Capacitance ($C$) vs Magnetic field ($B$) curves in the middle of the nematic range of each system were acquired to estimate permittivity values at infinite magnetic field through plotting ($1/\varepsilon'$) against $1/\mu_0 H$, according to the extrapolation method proposed by Clark et al.[56]

The extrapolated permittivity at infinite magnetic field was estimated 8-10% higher than the corresponding values of $\varepsilon_\parallel$ measured with $B = 1.4$ T. This difference arises from the unavoidable presence of two distorted layers close to the cell substrates, where the director is not uniformly aligned parallel to the magnetic field.[45]

**Table 1** Phase sequences and transition temperatures of CB9CB/5CB mixtures.

| Compound | 5CB wt% | Transition Temperatures (°C) |
|---|---|---|
| **CB9CB** | 0 | Iso 124.9 **N** 108.7 **$N_x$** |
| **DM9** | 9.1 | Iso 115.5 **N** 93.8 **$N_x$** |
| **DM21** | 21.4 | Iso 102.2 **N** 78 **$N_x$** |
| **DM37** | 36.8 | Iso 85.6 **N** 52.7 **$N_x$** |
| **DM51** | 51.4 | Iso 73.9 **N** 34.5 **$N_x$** |
| **5CB** | 100 | Iso 35.3 N |

At the onset of the N-$N_x$ phase transition, the magnetic field was not capable to reorient homeotropically the $N_x$ phase, thus *P.1* was followed only for the measurements within the N phase. For the determination of $\varepsilon^*_\parallel(\omega)$ in the $N_x$ phase a different procedure (*Protocol 2-P.2*) was used, based on the fact that upon the application of a sufficiently high electric field the $N_x$ phase can be irreversibly switched to a homeotropic state, which is stable unless the sample is heated back to the N phase.[18,29,57,58]

Specifically, the 20 μm planar samples were placed in the Novocontrol cryostat and slowly cooled at temperatures approximately 10°C below the Iso-N transition. At isothermal conditions, a 10 kHz sinusoidal voltage with $V_{rms} = 25$ V (E = 1.25 V/μm), which is well above the onset of the nematic Fréedericksz transition, was externally applied (TTi TGA1214 arbitrary waveform generator and HP6827A amplifier). In the presence of the external aligning electric field, samples were slowly cooled 40-45°C below the N-$N_x$ phase transition. Finally, the external aligning field was removed and $\varepsilon^*_\parallel(\omega)$ was measured upon heating using the 0.5 $V_{rms}$ probe field. To evaluate the induced homeotropic alignment during heating scans, samples were observed under the polarising microscope following the same steps as in the dielectric measurements (*P.2* protocol). A slow relaxation of the induced homeotropic alignment was observed close to the $N_x$-N transition in CB9CB, which was accelerated by the added presence of 5CB in the binary mixtures (see also Fig. SI2). It should be noted that homeotropic alignment under DC bias conditions was also tested, but not employed, since convective instabilities were optically observed, in agreement with previous studies.[30,58] In both experimental protocols, data acquisition and storage were controlled by WinDETA software from Novocontrol.

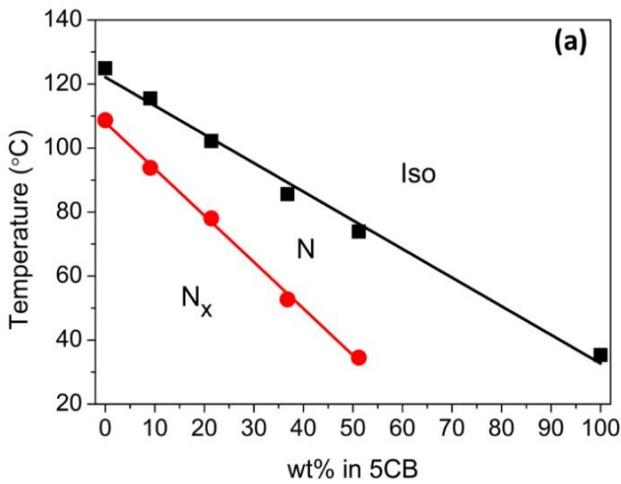

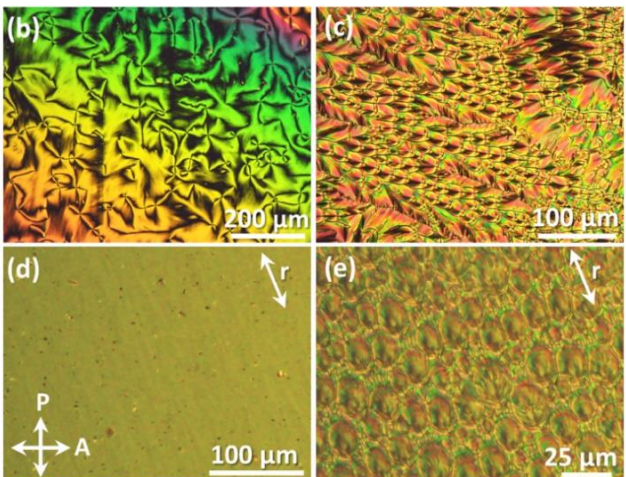

**Fig. 2** (a) Temperature-composition phase plot of CB9CB/5CB mixtures. Transition temperatures are defined through POM observations on cooling (10°C/min) in untreated glasses; The lines are a guide to the eye; Optical textures of mixture DM21 between crossed polarisers in untreated glasses: (b) N phase at 92.4°C, (c) $N_x$ phase at 48.4°C and in planar 20 μm cells: (d) N phase at 91.7 °C, (e) $N_x$ phase at 60 °C.

## Results and Discussion
### Mesomorphic Behaviour and Birefringence

Composition details and transition temperatures of the CB9CB/5CB (Dimer-Monomer, DM) mixtures are listed in Table 1. Transition temperatures, extracted from optical characterisation of the samples in untreated glasses, were used to construct the corresponding phase plot on cooling, presented in Fig. 2a. On increasing 5CB content a linear decrease of both the Iso-N and N- $N_x$ transition temperatures is observed, along with an expansion of the temperature range of the N phase. The linear dependence of transition temperatures with respect to concentration has been associated with ideal mixing of the constituents.[60] During optical characterisation, no cases of demixing were detected. However, on increasing the amount of 5CB a broadening of both the Iso-N and N-$N_x$ transitions is observed, a trend also reported in calorimetric studies of CB9CB/5CB[59] and CB7CB/5CB mixtures.[47,61] The emergence of the $N_x$ phase is detected up to mixture DM51. Nevertheless, DM51 exhibits an Iso-N phase

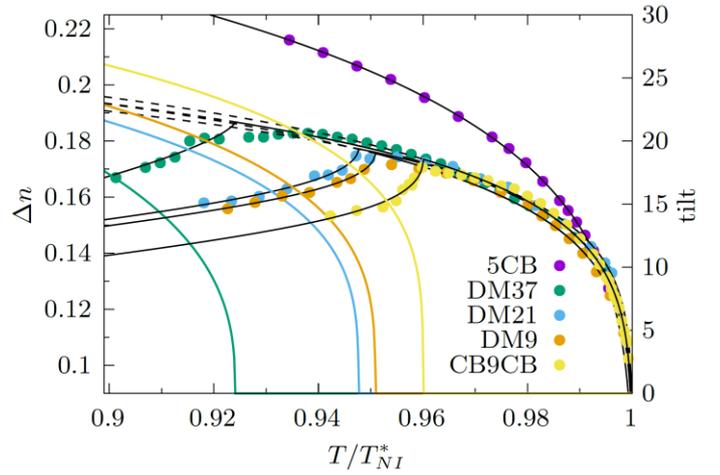

**Fig. 3** Left axis: Temperature dependence of the birefringence ($\Delta n$) in both nematic phases of CB9CB-5CB mixtures, together with the theoretical fitting (solid lines) according to Eqs. (1), (2). Dashed lines represent the extrapolation of fitted curves of the nematic phase. Right axis: Calculated (see Eq. 3) temperature dependence of the tilt angle ($\theta$) in the $N_x$ phase of all studied systems.

coexistence for a temperature window of about 12°C, followed by a very slow transition to the $N_x$ phase. Therefore, DM51 was not included in dielectric and birefringence experiments. The trends presented in the phase plot of Fig. 2 are consistent with earlier works considering cyanobiphenyl-based binary systems.[46,47,59,61,62]

Optical observations in untreated glasses during cooling from the isotropic phase, revealed typical textures of the N and $N_x$ phases. Representative examples are shown in Figs. 2b,c for mixture DM21. The N phase exhibits a schlieren texture, while in the $N_x$ phase rope-like configurations, as well as focal conic domains were developed. In treated cells for planar alignment (20 $\mu$m, antiparallel rubbing) the N phase is homogeneously aligned (Fig. 2d) and $N_x$ textures include the well-reported rope-like configurations (not shown here), however, the relatively thick cells promote the formation of focal-conic domains (Fig. 2e).

Fig. 3 depicts the temperature dependence of birefringence of all studied systems (left-axis) measured on cooling from the Isotropic phase. The measured birefringence, $\Delta n$, for all tested samples exhibits two discrete regimes, as illustrated in Fig.3. In the temperature range of the N phase, $\Delta n$ exhibits the typical temperature dependence of low molar mass uniaxial nematics, although some deviations are observed as the temperature approaches the N-$N_x$ phase transition. With the onset of the $N_x$ phase, $\Delta n$ drops abruptly and exhibits a much weaker decrease on lowering the temperature. The experimental data in the high temperature N phase are well fitted with the Haller equation:[40]

$$\Delta n(T) = \Delta n_0 \left(1 - \frac{T}{T_{IN}^*}\right)^b, \qquad T < T_{IN}^* \quad , \tag{1}$$

with fitting parameters $\Delta n_0, b$ and $T_{IN}^*$. $\Delta n_0$ corresponds to birefringence extrapolated to absolute zero, $T_{IN}^*$ to a temperature slightly above the Isotropic-Nematic transition temperature ($T_{IN}$) and $b$ to an exponent, which assumes values close to 0.2 for ordinary nematics.[40,63] Given the high chemical affinity between CB9CB and 5CB, the lower birefringence of pure CB9CB and its mixtures, compared to $\Delta n$ of the neat 5CB at the same reduced temperature, reflects mainly the lower global orientational order of DM mixtures. Interestingly, in the reduced temperature range of the N phase of CB9CB (up to $T/T_{IN}^* \approx 0.96$), birefringence is almost independent of 5CB concentration. This clearly suggests that the global orientational order of the mixtures is dictated by the ordering of CB9CB mesogens, therefore the orientational distribution of 5CB mesogens is adjusted to the corresponding distribution of the CB units of the dimer. Moreover, a deviation from the Haller behaviour of CB9CB close to the transition to the $N_x$ phase is observed, according to which the increase of $\Delta n$ is suspended and does not follow the theoretical prediction. This behaviour is enhanced in the mixtures and becomes very pronounced in DM37, where the N phase temperature range becomes significantly wider. Similar trends have been reported in birefringence[42,64,65] and NMR studies[9,66,67] of $N_x$ forming systems. This pronounced pretransitional behaviour suggests, that deeply in the N phase there is a substantial growth of the short range intermolecular orientational correlations that become long ranged with the onset of the transition to the low temperature nematic phase. The structure and the extend of the formed "clusters" depend strongly on 5CB concentration, as it is suggested by the widening of the pretransition range upon adding 5CB.

On entering the $N_x$ phase, birefringence decreases abruptly in both the dimer and the binary mixtures. This characteristic decrease of birefringence in the $N_x$ phase does not signify a decrease in the local orientational order, rather than a shift of the maximum of the orientational distribution of the cyanobiphenyl (CB) units with respect to the optic axis of the phase. The temperature dependence of birefringence within the $N_x$ phase can be described by a modified Haller equation:

$$\Delta n(T) = \Delta n_0 \left(1 - \frac{T}{T_{IN}^*}\right)^b P_2(\cos\theta(T)), \qquad T < T_{NN} \tag{2}$$

where $P_2(x) = (3x^2 - 1)/2$ is the second Legendre polynomial and

$$\theta(T) = a'\left(1 - \frac{T}{T_{NN}^*}\right)^{b'} \tag{3}$$

accounts for the temperature dependence of the deviation of the average direction of ordering of the CB units with respect to the optical axis of the mesophase. In Eq. (2), there are three new fitting parameters: $T_{NN}^*$ corresponding to the N-$N_x$ transition temperature, $a'$ the extrapolated "tilt" angle at absolute zero and $b'$ a critical exponent. A satisfactory fitting is achieved with Eq. (2) for the temperature range of measurements in the $N_x$ phase. The calculated values of $\theta(T)$ are presented in Fig. 3 (right axis) for neat CB9CB and DM mixtures. Fitting parameters for both nematic phases are listed in Table 2. Parameters $\Delta n_0, b$ extracted from the fitting in the N phase of both neat compounds are in good agreement to previous reported studies.[41,61,63] It should be noted, that birefringence measurements with the Berek compensator become less reliable deeper in the $N_x$ phase due to the development of the rope-like deformations, which cannot be avoided in the 5 $\mu$m cells used. For this reason, although the fitting represents satisfactorily the experimental results, extrapolation at lower temperatures should be treated with caution.

Considering the CB9CB dimer as two connected 5CB mesogens, the ratio $(\Delta n)_{CB9CB}/(\Delta n)_{5CB}$, at any specific reduced temperature, could be used as a rough measure of the ratio of the orientational order parameters associated with the ordering of the CB units. In the limit $T \to 0$, the shape of the dimer is expected to be close to its lowest energy conformation, and the above ratio can be used to estimate the effective bend angle, $\phi$, according to $\frac{(\Delta n_0)_{CB9CB}}{(\Delta n_0)_{5CB}} = \frac{1-3\cos^2(\phi)}{4}$.[41] Using for $\Delta n_0$ the values given in Table 2 for the neat CB9CB system, we get $\phi_{CB9CB} \approx 128°$, which is slightly higher from the $\phi_{CB7CB} \approx 122°$ bend angle of CB7CB,[41] estimated with the same method. The effective bend angle is within the range of the bend angles of the lowest energy configurations of cyanobiphenyl-based dimers connected with 9-atom flexible spacers, as they have been calculated with detailed molecular mechanics calculations.[68]

**Table 2** Optimal fitting parameters of the birefringence in N and $N_x$ phases of studied systems according to Eq. (2) and (3).

|  | $\Delta n_0$ | $b$ | $\alpha'$ | $b'$ | $T_{IN}$ | $T_{NN_x}$ |
|---|---|---|---|---|---|---|
| **5CB** | 0.368 | 0.196 | -- | -- | 36.4 | -- |
| **DM37** | 0.258 | 0.125 | 1.178 | 0.372 | 81.1 | 54.2 |
| **DM21** | 0.247 | 0.112 | 0.868 | 0.28 | 105.1 | 85.3 |
| **DM9** | 0.263 | 0.134 | 0.909 | 0.283 | 109.7 | 90.9 |
| **CB9CB** | 0.264 | 0.13 | 0.9 | 0.247 | 125.1 | 109.2 |

**Temperature dependence of static dielectric permittivity**

In the current work, dielectric measurements are performed in the frequency range between 100 Hz and 1 MHz. Typically, dielectric permittivity obtained at frequencies around 1-10 kHz is considered as static permittivity. This is, indeed, the case for both pure compounds, as well as for all binary mixtures in temperatures within the N phase since in this frequency window (1-10 kHz) there is no contribution of any relaxation mechanism. In the dielectric spectra acquired in DM mixtures, a relaxation mechanism ($m_1$) is recorded in the $N_x$ phase between $10^4$ -$10^6$ Hz, both in planar and homeotropic configurations. The low frequency relaxation ($m_1$) in symmetric CBnCB dimers [30,49] and in CB7CB/5CB mixtures[47] has been attributed to the flip-flop reorientations of the dipolar groups parallel to the director. Taking into account the existence of the relaxation mechanism above 10 kHz, as well as ionic conductivity effects in the low frequency region (below 1 kHz) of the recorded spectra (not included for brevity), the value of dielectric permittivity at 5 kHz is chosen as the most convenient to study the effect of the added content of 5CB on the dielectric anisotropy of binary systems.

The temperature dependence of $\varepsilon_\perp$ and $\varepsilon_\parallel$ components of the dielectric permittivity at 5 kHz is presented in Fig. 4 for all studied systems. Measurements of $\varepsilon_\parallel$ in the N phase of CB9CB and DM mixtures (full red circles) are performed on cooling in the presence of 1.4 T magnetic field following Protocol 1 (*P.1*). Open symbols correspond to $\varepsilon_\parallel$ (5 kHz) in the $N_x$ phase measured on heating without any aligning field following Protocol 2 (*P.2*).

The alignment within the $N_x$ phase was assessed through independent POM observations of the examined specimens according to *P.2*. Representative textures obtained for mixture DM21 can be found in Fig. SI1. During cooling from the N phase in the presence of the aligning field, the uniform homeotropic alignment is retained for a few degrees below the N-N$_x$ transition (Fig. SI1a). However, deep in the N$_x$ phase a distortion of the homeotropic texture is observed due to nucleation of toric focal conic domains and the subsequent growth of arrays of parabolic defects (Fig. SI1b). This is in agreement with previously reported observations in cyanobiphenyl dimers, [12,58,69] also reminiscent of the field-driven growth patterns observed in SmA phase.[70] Upon field removal, samples do not relax to the initial planar state and the observed patterns during subsequent heating (Fig SI1c,d) appear similar to those obtained on cooling with the electric field, following a reversible manner. Moreover, approaching the N$_x$-N transition the induced homeotropic alignment slowly relaxes back to planar in the CB9CB dimer. The temperature window, where loss of alignment is optically observed, becomes wider upon adding 5CB and as a result, measured values of $\varepsilon$ progressively decrease, reaching eventually the values measured with planar anchoring conditions in complete loss of the field-induced alignment. Measured permittivities at the corresponding temperature range (see Fig. SI2) are omitted from graphs of Fig.4 for clarity.

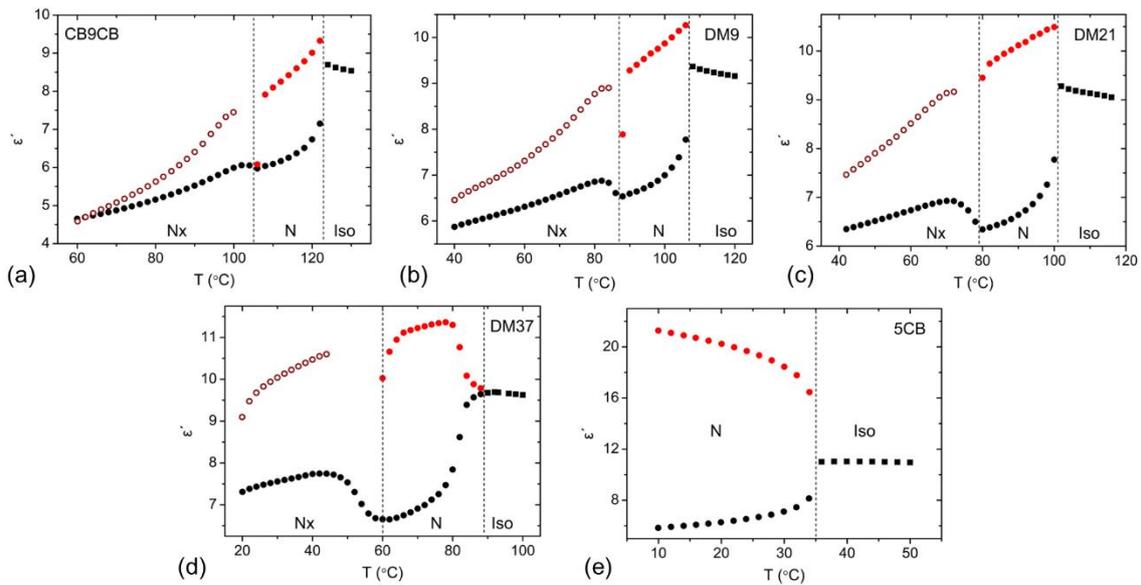

**Fig. 4** Temperature dependence of dielectric permittivity at f = 5 kHz of (a) CB9CB, (b) DM10, (c) DM21, (d) DM37 and (e) 5CB. Squares-$\varepsilon_{iso}$; Black circles-$\varepsilon_\perp$; Red circles–$\varepsilon_\parallel$ measured with *P.1*; Open circles–$\varepsilon_\parallel$ measured with *P.2*.

The temperature dependence of the components of the dielectric permittivity of CB9CB (Fig. 4a) and 5CB (Fig. 4e) is in reasonable agreement with previous studies.[30,45,71] Focusing on $\varepsilon_\parallel$, 5CB exhibits the typical temperature dependence observed in low molar mass nematogens with strong longitudinal dipole moment, while in CB9CB, $\varepsilon_\parallel$ increases at the IN transition and, subsequently, sharply decreases with decreasing temperature in both nematic phases, in agreement with previously reported results on odd CBnCB members.[18,30,43] Close to the N-N$_x$ phase transition the substantial development of short-range correlations results in the drastic reduction of $\varepsilon_\parallel$, especially in pure CB9CB and DM9 (Fig. 4a,b). At the onset of the transition to N$_x$, such correlations become long-range and the magnetic field can no longer retain the homeotropic alignment. The addition of 5CB inhibits the strong reduction of $\varepsilon_\parallel$ in CB9CB and entails the progressive increase of the dielectric anisotropy in both nematic phases of the DM systems (Fig. 4b-d). Interestingly, the pretransitional decrease of $\varepsilon_\parallel$ deep in the N phase smoothens with increasing 5CB content, especially in DM37 (Fig. 4d). We should recall, that the N-N$_x$ transition becomes progressively weaker with the addition of 5CB.[47,59] These results suggest that the extent of the short-range correlations is dependent on 5CB content, in agreement to our birefringence measurements, and at high concentrations of the monomer become continuously long-ranged. In the following theoretical description, the data close to the N-N$_x$ transition have not been considered.

The most prominent feature of the dielectric behaviour of DM mixtures is the temperature dependence of $\varepsilon_\perp$ in the N$_x$ phase, in which the addition of 5CB seems to have a greater impact. This is reflected by the formation of a hump right after the N- N$_x$ transition in DM9. This hump is significantly enhanced in DM21, while in DM37 $\varepsilon_\perp$ increases after the N-N$_x$ transition and then is almost stabilised. The observed trends are in agreement with previous studies on the dielectric anisotropy of CB7CB based dimer-monomer binary systems.[46,47]

**Kirkwood-Fröhlich Theory for Flexible Dimers and Dimer-Monomer Mixtures**

In this section, and in order to reproduce the experimental results obtained for the dielectric permittivities of all systems studied in this work, a molecular field theory in conjunction with the Kirkwood-Fröhlich theory for dielectric permittivity of anisotropic polar liquids is introduced for the neat compounds and mixtures.

According to the anisotropic version of the Kirkwood-Fröhlich equation[72,73] for polar fluids, the dielectric constant $\varepsilon_\lambda$ along a certain direction $\hat{\boldsymbol{\lambda}}$ of a uniform anisotropic material at temperature $T$, is connected to the molecular dipole moment $\boldsymbol{m}$ and the total polarization $\boldsymbol{M} = \sum_i \boldsymbol{m}_i$ of a spherical volume that contains $N$ molecules and is embedded in the medium, according to

$$\frac{(\varepsilon_\lambda - \varepsilon_{\infty\lambda})(\varepsilon_\lambda + (\varepsilon_{\infty\lambda} - \varepsilon_\lambda)\Omega_\lambda)}{\varepsilon_\lambda(\varepsilon_{\infty\lambda} + 2)^2} = \frac{\rho}{9\varepsilon_0 k_B T} \langle (\boldsymbol{m} \cdot \widehat{\boldsymbol{\lambda}})(\boldsymbol{M} \cdot \widehat{\boldsymbol{\lambda}}) \rangle \tag{4}$$

where $\rho = N/V$ is the molecular number density, $\varepsilon_0$ is the vacuum permittivity, $k_B$ is the Boltzmann constant, $\Omega_\lambda$ is the factor that accounts for the depolarization field associated with a sphere in an anisotropic medium, taken as $1/3$ in this work,[72] and $\varepsilon_{\infty\lambda}$ the high frequency limit of the dielectric constant along the $\lambda$-principal axis. The brackets denote equilibrium averaging over molecular positions, orientations and, in the case of flexible molecules, over molecular conformations.

The dielectric constant in the optical regime is related to the refractive index of the material as $\varepsilon_{\infty,\lambda} = n_\lambda^2$. Assuming that the temperature dependence of the refractive index is qualitatively described by $n_\parallel = n_{iso} + 2\Delta n/3$ and $n_\perp = n_{iso} - \Delta n/3$, with $\Delta n$ the temperature depended birefringence of the nematic and $n_{iso}$ the refractive index of the isotropic phase, which is assumed constant at least at temperatures close to $T_{NI}$.

Setting $[\mu^2]_\lambda \equiv \langle (\boldsymbol{m} \cdot \widehat{\boldsymbol{\lambda}})(\boldsymbol{M} \cdot \widehat{\boldsymbol{\lambda}}) \rangle$ and working in the director ($\widehat{\boldsymbol{n}}$) frame of a macroscopically uniaxial fluid, we get:

$$[\mu^2]_{\parallel(\perp)} = m_{\parallel(\perp)}^2 g_{1,\parallel(\perp)}^{(inter)} \tag{5}$$

with $m_\parallel^2 = \langle (\boldsymbol{m} \cdot \widehat{\boldsymbol{n}})^2 \rangle$ and $2m_\perp^2 = \langle \boldsymbol{m}^2 \rangle - m_\parallel^2$ the mean squared projections of the *total* molecular dipole, $\boldsymbol{m}$, along and perpendicular to the director respectively, and

$$\begin{aligned} g_{1,\parallel}^{(inter)} &= 1 + \rho \frac{\langle (\boldsymbol{m} \cdot \widehat{\boldsymbol{n}})(\boldsymbol{m}' \cdot \widehat{\boldsymbol{n}}) \rangle}{m_\parallel^2} \\ g_{1,\perp}^{(inter)} &= 1 + \rho \left( \frac{\langle \boldsymbol{m} \cdot \boldsymbol{m}' \rangle - \langle (\boldsymbol{m} \cdot \widehat{\boldsymbol{n}})(\boldsymbol{m}' \cdot \widehat{\boldsymbol{n}}) \rangle}{2m_\perp^2} \right) \end{aligned} \tag{6}$$

are the respective Kirkwood *inter*-molecular dipole-dipole correlation factors. Clearly, $[\mu^2]_\lambda$ gives the mean square of the *effective* total molecular dipole moment along the direction $\widehat{\boldsymbol{\lambda}}$ and $g_{1,\parallel(\perp)}^{(inter)}$ measures the extent of orientational correlations between the projections of the total dipoles of different molecules along and perpendicular to the symmetry axis of the phase. In the isotropic phase, for any direction $\widehat{\boldsymbol{\lambda}}$, we get $m_\lambda^2 = \langle \boldsymbol{m}^2 \rangle/3$, $\langle (\boldsymbol{m} \cdot \widehat{\boldsymbol{\lambda}})(\boldsymbol{m}' \cdot \widehat{\boldsymbol{\lambda}}) \rangle = \langle \boldsymbol{m} \cdot \boldsymbol{m}' \rangle/3$ and $g_{1,\lambda}^{(inter)} = 1 + \rho \frac{\langle \boldsymbol{m} \cdot \boldsymbol{m}' \rangle}{\langle \boldsymbol{m}^2 \rangle}$.

In the case of flexible particles having more than one dipolar groups which are not rigidly connected, $\boldsymbol{m}$ and $\boldsymbol{m}'$ correspond to the conformational dependent *total* dipole moment of different molecules. Each of the molecular dipolar segments may make separate contributions to the dielectric constant. We note here that the correlation factors $g_{1,\parallel(\perp)}^{(inter)}$ in Eqs (6) do not contain contributions of the intramolecular dipolar correlations, since this specific information is integrated in the conformationally averaged $m_{\parallel(\perp)}^2$ terms. To separate these contributions, we take into account that the net molecular dipole moment due to the permanent dipoles, $\boldsymbol{m}$, of a given molecular conformation having the cyanobiphenyl units pointing in the $\widehat{\boldsymbol{L}}_1$ and $\widehat{\boldsymbol{L}}_2$ directions, is given by $\boldsymbol{m} = \mu_{CN}(\widehat{\boldsymbol{L}}_1 + \widehat{\boldsymbol{L}}_2)$ with, $\mu_{CN}$, the electric dipole moment of the CN terminal group. In this case, we can rewrite the mean-square projections of the total molecular dipole moment as:

$$\begin{aligned} (m_\parallel^2)_{dim} &= 2\mu_{CN}^2 \frac{1}{3}(1 + 2S_d) g_{1,\parallel}^{(intra)}, \quad g_{1,\parallel}^{(intra)} = 1 + 3\frac{\langle (\widehat{\boldsymbol{L}}_1 \cdot \widehat{\boldsymbol{n}})(\widehat{\boldsymbol{L}}_2 \cdot \widehat{\boldsymbol{n}}) \rangle_b}{1 + 2S_d} \\ (m_\perp^2)_{dim} &= 2\mu_{CN}^2 \frac{1}{3}(1 - S_d) g_{1,\perp}^{(intra)}, \quad g_{1,\perp}^{(intra)} = 1 + \frac{3}{2}\frac{\langle \widehat{\boldsymbol{L}}_1 \cdot \widehat{\boldsymbol{L}}_2 \rangle_b - \langle (\widehat{\boldsymbol{L}}_1 \cdot \widehat{\boldsymbol{n}})(\widehat{\boldsymbol{L}}_2 \cdot \widehat{\boldsymbol{n}}) \rangle_b}{1 - S_d} \end{aligned} \tag{7}$$

with $S_d = \langle P_2(\hat{L}_1 \cdot \hat{n}) + P_2(\hat{L}_2 \cdot \hat{n}) \rangle / 2$ the nematic orientational order parameter associated with the ordering of the CB units of the dimer (the derivation of Eqs (7) is given in SI). The subscript $(b)$ in the averages defining the Kirkwood factors, is used to remind that the orientations $\hat{L}_{1(2)}$ are referred exclusively to the "bonded" CB units of a single dimer. Consequently, $g_{1,\|(\perp)}^{(intra)}$ are direct measures of the contribution of the *intra*-molecular dipolar correlations to the total Kirkwood factors in Eq (5).

In agreement with previous theoretical studies,[49,50] Eqs (7) demonstrate clearly that the intrinsic intramolecular dipolar correlations, may change substantially the dielectric permittivity of the dimers with respect to those of the monomeric systems, not only in the nematic, but also in the isotropic phase where $g_{1,iso}^{(intra)} = 1 + \langle \hat{L}_1 \cdot \hat{L}_2 \rangle_b$. In the case of the 5CB mesogens (the "monomer"), we have $g_1^{(intra)} = 1$ and Eqs (7) reduce to the familiar relations for rigid dipolar rod-like uniaxial particles:[72]

$$(m_\|^2)_{mon} = \mu_{CN}^2(1 + 2S_m)/3,$$
$$(m_\perp^2)_{mon} = \mu_{CN}^2(1 - S_m)/3 \tag{8}$$

In the case of a homogeneous mixture of $N_m$ monomers with $N_d$ dimers, Eq. 5 is written as $[\mu^2]_\lambda \equiv (\chi_d(m_\lambda^2)_{dim} + \chi_m(m_\lambda^2)_{mon})g_{1,\lambda}^{(inter)}$, where $\chi_{m(d)} = N_{m(d)}/(N_m + N_d)$ is the mole fraction of the mixture, $\lambda = \|, \perp$ denotes the direction along or perpendicular to the director and $g_{1,\lambda}^{(inter)}$ the Kirkwood corelation factor between dipoles belonging to different molecules of the same or different species. Inserting the definitions for $(m_\lambda^2)_{dim(mon)}$, we get:

$$[\mu^2]_\| \equiv \mu_{CN}^2 \frac{1}{3}\left(2\chi_d(1 + 2S_d)g_{1,\|}^{(intra)} + \chi_m(1 + 2S_m)\right)g_{1,\|}^{(inter)}$$

$$[\mu^2]_\perp \equiv \mu_{CN}^2 \frac{1}{3}\left(2\chi_d(1 - S_d)g_{1,\perp}^{(intra)} + \chi_m(1 - S_m)\right)g_{1,\perp}^{(inter)} \tag{9}$$

Equations (9) reduce to the corresponding equations derived for the neat compounds when $\chi_{m(d)} = 1$.

To estimate theoretically the temperature dependence of the order parameters $S_{d(m)}$ and of the *intra*-molecular correlation factors $g_{1,\lambda}^{(intra)}$, we introduce a Maier-Saupe (MS) like, mean field (MF) theory for mixtures of CB9CB (the symmetric dimer) with 5CB (the monomer). To do this, we consider a uniform mixture of $N_d$ dimers with $N_m$ monomers at constant volume and temperature $T$. In the presence of a uniaxial nematic field, we assume that the CB groups of the mesogens experience an ordering potential of the form $u_{CB}(\hat{L}) = -wP_2(\hat{n} \cdot \hat{L})$, where $P_2$ is the second Legendre polynomial of the projection of direction $\hat{L}$ of the CB-unit on the nematic director $\hat{n}$ and $w$ is an energy parameter, that depends on the specific thermodynamic conditions and is associated with the aligning strength of the nematic field. Note here that we do not consider explicitly the orientational couplings between the end-chain of 5CB, nor of the flexible spacer of the dimer with the nematic director, which would result in an improvement to the description of the potential of mean torque of flexible mesogens,[34] in order to keep the model as simple as possible. With these assumptions the MF potential of the monomer and of the dimer can be written in a common general form as $U_{MF}^{(s)}(\omega, n_s) = wW^{(s)}(\omega, n_s)$ with the ordering function $W^{(s)}(\omega, n_s)$ representing the form of the orientational coupling of the s-type mesogen, when it is at its $n_s$ conformational state with orientation $(\omega)$ with respect to the nematic director. Consequently, the ordering function of the 5CB monomer is $W^{(m)}(\hat{L}) = P_2(\hat{n} \cdot \hat{L})$ and of the symmetric dimer, when it is at its $n_s$ conformational state, is taken to be of the form $W^{(d)}(\hat{L}_1, \hat{L}_2; n_s) = P_2(\hat{n} \cdot \hat{L}_1) + P_2(\hat{n} \cdot \hat{L}_2) + b|\hat{L}_1 \times \hat{L}_2|P_2(\hat{n} \cdot \hat{s})$; with $\hat{L}_{1(2)}$ the directions of the CB groups at the given molecular conformation $n_s$, and $\hat{s} = (\hat{L}_1 \times \hat{L}_2)/|\hat{L}_1 \times \hat{L}_2|$ the unit vector normal to the plane defined by the directions of the two CB groups. The first two terms in $W^{(d)}$ are the dominant ones that describe the ordering of the CB-units of the dimer, while the last term is included for completeness and takes into account the inherent molecular biaxiality of the dimer, with $b$ being a parameter quantifying the relative contribution of the biaxial term to the overall ordering. Any value $b < 0.1$ could be used without significant changes to the obtained results; here we have used $b = 0.05$ corresponding to a small biaxial perturbation that reproduces the NMR obtained jump of the principal order parameter $(S_d)_{NI}$ of the neat CB9CB at the NI phase transition.[9]

With these conventions the conformation-dependent orientational probability distribution function of the $s = d, m$ molecular species is of the form:

$$f^{(s)}(\omega, n_s) = p_{n_s}^{(s)} exp[-U_{MF}^{(s)}(\omega, n_s)/k_B T]/\zeta^{(s)} \quad (10)$$

with $p_{n_s}^{(s)}$ the statistical weight of the $n_s$ molecular conformation of a single molecule in the isotropic phase, associated with the *intra*molecular potential energy $E^{(s)}(n_s)$ according to $p_{n_s}^{(s)} = exp[-E^{(s)}(n_s)/k_B T]/\zeta_{conf}^{(s)}$, with $\zeta_{conf}^{(s)} = \sum_{n_s} exp[-E^{(s)}(n_s)/k_B T]$ the conformational molecular partition function. The normalization constant in Eq. (10) is given by $\zeta^{(s)} = \sum_{n_s} p_{n_s}^{(s)} \zeta_{n_s}^{(s)}$ with $\zeta_{n_s}^{(s)} = \int d\omega \, exp[-U_{MF}^{(s)}(\omega, n_s)/k_B T]$. With these definitions, the average of any conformational/orientational dependent single molecule property $A(\omega, n)$ is calculated as $\langle A^{(s)}(\omega, n) \rangle = \sum_n \int d\omega f^{(s)}(\omega, n) A^{(s)}(\omega, n)$.

Assuming a weak first order phase transition, at which the coexisting phases have the same density and composition, the free energy difference of a homogeneously ordered system ($\langle W^{(s)} \rangle \neq 0$) with respect to the free energy of the isotropic phase ($\langle W^{(s)} \rangle = 0$) at the same thermodynamic conditions, is given by:

$$\frac{\Delta F}{V k_B T} = -\sum_s \rho \chi^{(s)} \left( ln(\zeta^{(s)}/\zeta_{iso}^{(s)}) + \frac{w}{2k_B T} \langle W^{(s)} \rangle \right) \quad (11)$$

We note here that once the conformational properties of the isolated molecules are known, the free energy of the system at constant composition and density depends only on the dimensionless ratio $w/k_B T$. Furthermore, on the mean field level, the free energy is minimized with respect to the variational energy parameter $w$, when the latter is of the form $w/k_B T = -w_0 \sum_s \chi^{(s)} \langle W^{(s)}(\omega, n) \rangle_s$ where $w_0$ is a parameter dependent of the composition and density of the system.

There are several methods, at different levels of resolution, to model the accessible molecular conformations of the flexible dimer and to calculate the distribution of the angles between the dipolar cores of the isolated dimer.[49,50,68] Here, we use a simple scheme according to which the (orientationally invariant) effective internal conformational energy of the single dimer depends on $c \equiv \hat{L}_1 \cdot \hat{L}_2$ as $E(c) = \sum_{i=1}^{3} a_i P_i(c)$, with $P_i(c)$ the $i$-rank Legendre polynomial. The $a_i$ parameters are adjusted to mimic the dominant features of the conformational statistics of the CB9CB dimer,[49,50,68] and to reproduce the NI phase transition temperature of the CB9CB. The dominance of the first three terms on the behaviour of the orientational coupling of the bonded mesogens is an intrinsic property of these dimers and not the circumstantial outcome of a particular modelling. With this continuous conformation model the conformational/orientational dependent probability distribution for the dimer given in Eq (10) can be written as $f^{(d)}(\omega, c)$ and the summation over the molecular conformations is replaced by integration over $c$ in the range $-1 < c < 1$.

According to the MS theory, at the NI phase transition we have $w_0/k_B T_{NI} \approx 4.54$. Therefore, given that the NI phase transition of neat 5CB system at $(T_{NI})_{5CB} = 308K$ we get that $w_0/k_B \approx 1398K$. Taking into account that $(T_{NI})_{CB9CB} = 393K$, we have chosen the set of the $a_l$ parameters as $a_1 = 0.25w_0$, $a_2 = 0.6w_0$ and $a_3 = -0.35w_0$ to reproduce the NI transition temperature of the neat CB9CB system. Details on the conformational statistics of the CB9CB dimer, as well as on the temperature dependence of the order parameter and of several intramolecular orientational correlation factors are given in Section II of the SI.

To calculate the temperature variation of the mean-square projections (Eqs (4) and (5)), of the total molecular dipole in the low temperature nematic phase ($T < T_{NN}$), we proceed utilizing the experimentally verified facts, that the $N_x$ phase of the pure dimer and of its mixtures correspond to a 1D modulated nematic fluid with heliconical structure having periodicity in the nanometre length-scale. In this picture, the nanoscale periodic modulation corresponds to a "hard" ordering mode, as opposed to the "soft", elasticity driven, director deformations in nematics, that occur in the micron-scale.[28] Because of the periodicity and the form of the 1D molecular order modulation, any second rank tensorial property, statistically averaged within a "small" spherical volume of the sample with radius equal or a few multiples of the pitch, has the form of a uniaxial tensor with symmetry axis the axis of modulation, say $\hat{\boldsymbol{h}}$. This spatial averaging eliminates the local polarity, an intrinsic feature of the $N_x$ phase. Therefore, Equations (4)-(9), derived for a uniaxial medium, hold also in the low temperature nematic phase provided that the direction $\hat{\boldsymbol{n}}$ is replaced by the direction of the axis of modulation $\hat{\boldsymbol{h}}$. Accordingly, all the involved averages, as well as the directions $\lambda = \|, \perp$ in the $N_x$ phase, are referred with respect to $\hat{\boldsymbol{h}}$. We note here that, as in the case of the conventional nematic phase, the theoretical treatment concerning the $N_x$ phase is referred to states with uniform alignment of the h-director.

Working in a lab frame with the macroscopic z-axis coinciding with $\hat{\boldsymbol{h}}$, we denote with $\hat{\boldsymbol{n}}_h(z)$ the *local* axis of the maximal alignment of the CB units at z. Assuming that $\hat{\boldsymbol{n}}_h(z)$ twists along $\hat{\boldsymbol{h}}$ in a heliconical fashion with conical angle $\theta$, after uniform integration over a helical pitch, we obtain the relationship between the order parameters in the laboratory frame and of the local ones. For instance, ignoring biaxiality, the scalar orientational order parameter of the CB units in the h-system reads $\langle P_2(\hat{\boldsymbol{L}} \cdot \hat{\boldsymbol{h}}) \rangle_{m(d)} = S_{m(d)} P_2(\cos\theta)$, with $S_{m(d)}$ the uniaxial order parameter in the local frame. Similarly, the average involved in the calculation of the intramolecular dipole-dipole correlations in the $h$-frame reads, $3\langle (\hat{\boldsymbol{L}}_1 \cdot \hat{\boldsymbol{h}})(\hat{\boldsymbol{L}}_2 \cdot \hat{\boldsymbol{h}}) \rangle_b - \langle \hat{\boldsymbol{L}}_1 \cdot \hat{\boldsymbol{L}}_2 \rangle = \left( 3\langle (\hat{\boldsymbol{L}}_1 \cdot \hat{\boldsymbol{n}}_h)(\hat{\boldsymbol{L}}_2 \cdot \hat{\boldsymbol{n}}_h) \rangle_b - \langle \hat{\boldsymbol{L}}_1 \cdot \hat{\boldsymbol{L}}_2 \rangle \right) P_2(\cos\theta)$.

The N-$N_x$ phase transition temperature and the temperature dependence of the "tilt" angle of the principal ordering axis of the CB units, assumed to be approximately the same for monomers and dimers, in the low temperature nematic phase is obtained directly by the birefringence measurements. The corresponding temperature dependence of the local ordering properties are calculated from the MF theory, as if the system was in its uniform nematic phase. This approximation is well justified from the successful fitting of the birefringence.

Assuming that the mass density is approximately the same for all studied systems and taking into account, that the molecular weight of the dimer, $M_{CB9CB}$, is practically twice of that of the monomer, $M_{CB9CB} \approx 2M_{5CB}$, it turns out that the number density, $\rho_{CN}$, of the CN groups is constant and independent of the composition of the mixture. Assuming a mass density of $\sim 1.0 \text{g/cm}^3$ in the whole temperature range, we get that $\rho_{CN}/9\varepsilon_0 k_B \approx 220 \text{ K/D}^2$. The refractive indices $n_{\|} = n_{iso} + 2\Delta n/3$ and $n_\perp = n_{iso} - \Delta n/3$ are estimated using the experimentally calculated temperature dependence of $\Delta n$ for each sample (shown in Fig. 3), with $n_{iso} = 1.6$ for all systems, since this is a typical value for nCBs[74] and CBnCBs.[44]

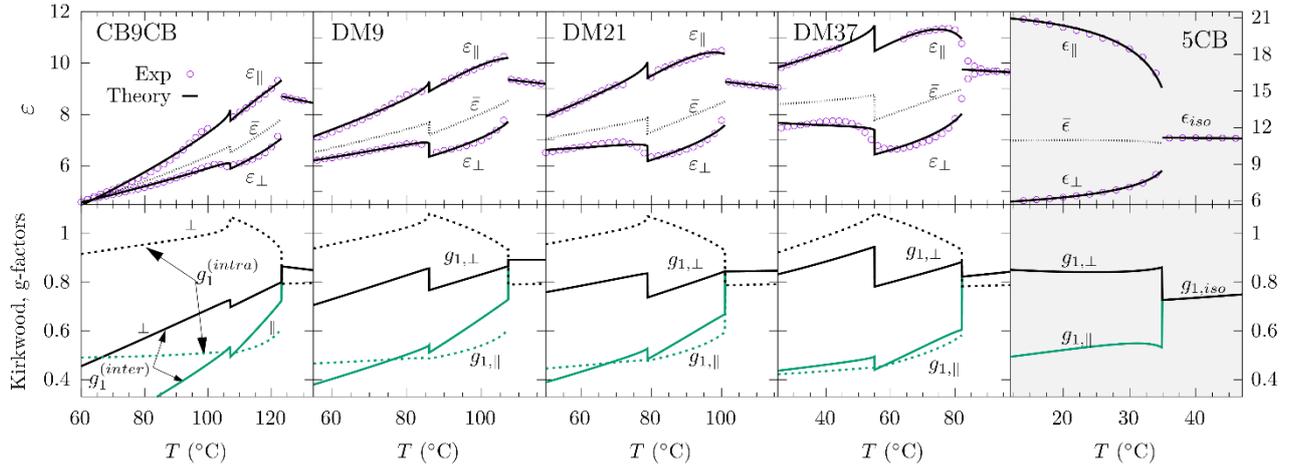

**Fig.5** (Top) Temperature dependence of the components of the static dielectric permittivity of the neat CB9CB, 5CB and DM mixtures. Open circles correspond to experimental measurements and solid lines to theoretical predictions. (Bottom) Kirkwood correlation factors in the Isotropic and Nematic phases. *Inter*-molecular dipole-dipole correlations, $g_{1,\|(\perp)}^{(inter)}$, are presented with the solid lines and the *intra*-molecular, $g_{1,\|(\perp)}^{(intra)}$, with dashed lines, with the black and green coloured lines corresponding to the perpendicular ($\perp$) and parallel ($\|$) components, respectively. In the neat 5CB system $g_{1,\|(\perp)}$, are referred exclusively to *inter*-molecular dipole-dipole corelations.

**Dipole-dipole associations and temperature/composition dependence of dielectric permittivity**

Up to this point, we have introduced a minimalistic model that satisfies the $C_{2V}$ symmetry of a symmetric dimer and captures its key conformational properties in an isotropic environment. In addition, the experimentally measured phase transition temperatures of the studied systems, have been utilised to build a parameter-free M-S mean field theory for the self-consistent calculation of the temperature/concentration dependence of the orientational order parameters related to the ordering of the CB-groups, as well as the calculation of the statistics of conformations of the dimer in an anisotropic environment and the corresponding intramolecular orientational correlations. The required tilt to describe the nanoscale modulation of the orientational order in the $N_x$ phase is obtained directly from birefringence measurements and is assumed to be the same for CB units of the dimer and of the monomers in the mixtures. The only term that enters in the Kirkwood-Fröhlich theory of dielectric permittivity and it is not possible to be calculated in the MF level are the orientational correlations between the CN dipoles belonging to different molecules. To take these correlations into account, we assume that the temperature dependence of the Kirkwood correlation factors can be expanded in a power series of the reduced temperature and represented as $g_{1,\|(\perp)}^{(inter)} = 1 + \alpha_{\|(\perp)}x + \beta_{\|(\perp)}x^2 + O(x^3)$, $x = T/T_{NI}$, and as $g_{1,iso}^{(inter)} = 1 + \alpha_0 x + \beta_0 x^2 + O(x^3)$, in the nematic and the isotropic phases respectively. Truncating the power series at the 3$^{rd}$ rank terms, the coefficients $\alpha_w, \beta_w, w = 0, \|, \perp$ are the only adjustable parameters, different for each phase, required to fit the experimentally measured static dielectric permittivity in the whole temperature range.

In the top graphs of Fig. 5, we present the experimental results (open circles) together with the theoretical predictions (solid lines) of the temperature dependence of the dielectric permittivity in the isotropic and nematic phases of all studied systems. The corresponding dipole correlation factors are presented in the bottom graphs, where the solid and dashed lines refer to the *inter-* and *intra*-molecular correlations, respectively. Note that for the 5CB system, a grey background is used to indicate the different scale of the $\varepsilon$-axis (right axis in Fig. 5).

In view of the several approximations already involved in the Kirkwood-Fröhlich theory for the dielectric permittivity, and of the rough estimations of the values of the prefactors of the mean square dipole moments, the success of the MF approximation is rather satisfactory provided that corrections due to the *inter*-molecular dipolar correlations are considered. More importantly, ignoring these correlations results to complete failure to reproduce the experimental measurements even in qualitative grounds.

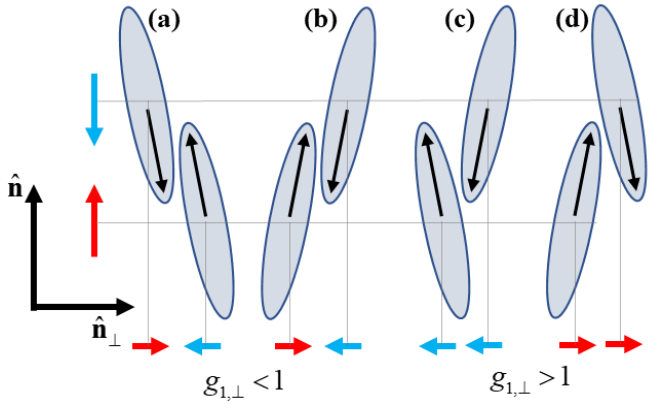

**Fig. 6.** Four different possible pair arrangements of rigid, rodlike, dipolar particles. In the MF approximation all have the same probability, since the orientations of the individual particles are either $\phi$ or $\pi - \phi$, where $\phi$ is the angle between the molecular direction $\hat{L}$ (arrows within the particles) and the symmetry axis of the phase, $\hat{n}$. $(\hat{L}_1 \cdot \hat{n})(\hat{L}_2 \cdot \hat{n})$ is the same for all the pairs, therefore contributing identically to $g_{1,\parallel}^{(inter)}$. However, only pairs with synclinic, with respect to the director, orientations, i.e. the pairs (a) and (b), have $(\hat{L}_1 \cdot \hat{n}_\perp)(\hat{L}_2 \cdot \hat{n}_\perp) < 0$ and therefore contribute with higher probability to $g_{1,\perp} < 1$, suggesting a discrimination between synclinic and anticlinic antiparallel arrangements in favour of the former.

**Neat 5CB system**

To facilitate the rationalization of our experimental findings for the neat dimer and its mixtures, we discuss initially in some detail our results on the simpler and extensively studied neat 5CB system. The temperature dependence of the Kirkwood dipolar factors of 5CB, suggest that the antiparallel association of the molecular dipoles, already present in the isotropic phase, exhibits a strong variation at the NI phase transition. As it is suggested by the theory, the jump of $\Delta \varepsilon = \varepsilon_\parallel - \varepsilon_\perp$ at the NI phase transition, as well as its temperature dependence in the nematic phase, are associated with two distinct mechanisms. The first mechanism is connected with the onset of long range orientational order, a collective property that alters the mean square dipole moment $[\mu^2]_{iso}$ of the isotropic phase by a multiplication factor $1 + 2S$ and a diminishing factor $1 - S$ in directions parallel and perpendicular to the nematic director, respectively (see Eq. (8)). In the case of 5CB, the onset of the long range orientational order is not sufficient to capture quantitatively neither the jump of $\Delta\varepsilon$, nor its temperature dependence in the nematic phase. This discrepancy is relaxed considering the second mechanism that is associated with the development of local intermolecular orientational dipolar correlations along and perpendicular to the nematic director. These correlations reduce significantly $[\mu^2]_\parallel$ due to substantial antiparallel correlations between the projections of the molecular dipoles of neighbouring mesogens on the nematic director. Interestingly, the fact that $g_{1,\perp} < 1$ suggests, that antiparallel correlations, to a lesser extent, are also present between the projections of the dipole moments perpendicular to the nematic director. This finding indicates a preference towards states, where neighbouring molecules rotate in a synchronized fashion, i.e., as a single supramolecular antiparallel associated entity, as illustrated and discussed in Fig. 6. We note here that, comparable values for the g-factors in the nematic phase of 5CB have been previously reported (at a single temperature) on the basis of Kirkwood-Fröhlich approximations.[53]

**Neat CB9CB and DM mixtures**

Concerning the static dielectric properties of the neat CB9CB system, we note initially that its dielectric permittivity in the isotropic phase exhibits stronger temperature variation compared to the corresponding variation of its monomeric analogue, as evident from Fig. 4 and Fig. 5. This is in accordance with previous studies on CBnCB systems.[43,44,46,47] At temperatures slightly above the NI phase transition the dielectric permittivities of the neat compounds are $\varepsilon_{5CB} \approx 11$ (see Ref. 45 and references therein) and $\varepsilon_{CB9CB} \approx 8.5$. Assuming that at the corresponding temperatures both materials have similar $\varepsilon_\infty$, for the ratios of both sides of Eq (4) we obtain $\left(\frac{\mu_{CN}^2 \rho_{5CB}}{2\mu_{CN}^2 \rho_{CB9CB}} \frac{T_{NI;CB9CB}}{T_{NI;5CB}} \frac{g_{1;iso;5CB}}{g_{1,iso;CB9CB}}\right) \simeq 1.3$, with $g_{1,iso}$ the dipole correlation factor between the dipole moments of the CN groups in the isotropic phase of the dimer and of the monomer. Substituting the measured transition temperatures and taking into account, that both materials have approximately the same mass density, which gives that $\rho_{5CB} \approx 2\rho_{CB9CB}$, we conclude that $g_{1,iso;5CB} \simeq g_{1,iso;CB9CB}$. In other words, both materials have similar Kirkwood dipole correlation factors at temperatures just above the onset of the NI phase transition. This observation becomes rather interesting given that the onset of the nematic phase results to a quantitatively and qualitatively different dielectric behaviour of the two materials. Since the g-factor of 5CB is connected exclusively with intermolecular dipole-dipole orientational correlations, while the corresponding g-factor of the symmetric dimer accounts for both *intra*- and *inter*-molecular correlations, the lower $\varepsilon_{iso}$ of CB9CB compared to that of 5CB can be attributed to a slightly higher population of the more elongated molecular conformers. Otherwise, in the absence of any internal orientational correlations CB9CB should have similar $\varepsilon_{iso}$ to 5CB.

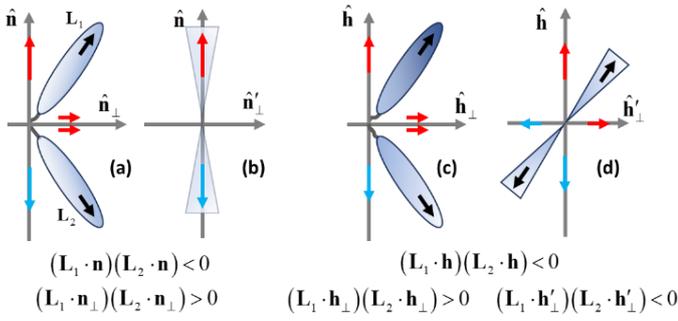

**Fig. 7** *Intra*-molecular dipolar correlations of the symmetric dimers in the N and in the $N_x$ phases. An elongated bent-shaped molecular conformation is assumed for the demonstration. The direction of the CN dipoles in the two arms are denoted as $L_1$ and $L_2$ and their projections to the axis of the director frame are represented with coloured arrows. In N phase, (a) and (b), the projections of the molecular dipoles on the director $\mathbf{n}$ leads to negative values of $(L_1 \cdot \mathbf{n})(L_2 \cdot \mathbf{n})$, while in directions perpendicular to $\mathbf{n}$ we have $(L_1 \cdot \mathbf{n}_\perp)(L_2 \cdot \mathbf{n}_\perp) \geq 0$ (see Fig. SI5). An increment in the bend angle of the dimer has opposite effects on the magnitudes of the two correlations, i.e., the magnitude of the correlation along $\mathbf{n}$ increases, while the corresponding magnitude along $\mathbf{n}_\perp$ diminishes towards zero, as the bend angle increases (lowering the temperature). Accordingly, in the nematic phase we have $g_{1,\perp,\parallel}^{(intra)} <$ 1. In the $N_x$ phase, the plane defined by the two arms of the dimer does not contain the symmetry axis of the phase, $\mathbf{h}$. At any given tilt, we have $(L_1 \cdot \mathbf{h})(L_2 \cdot \mathbf{h}) < 0$, but in directions perpendicular to $\mathbf{h}$ we have either positive correlations (c) or negative (d). The total correlation perpendicular to $\mathbf{h}$ is obtained from the mean value $((L_1 \cdot \mathbf{h}_\perp)(L_2 \cdot \mathbf{h}_\perp) + (L_1 \cdot \mathbf{h}'_\perp)(L_2 \cdot \mathbf{h}'_\perp))/2$, which for small tilts is positive and may change sign as the tilt increases. As the bend angle increases, the sign reversal takes place at smaller tilt angles. Due to these variations $g_{1,\perp}^{(intra)}$ is greater than unity with the onset of the $N_x$ phase and shifts to values lower than unity on lowering the temperature.

*Intra*-molecular orientational correlations

The bonded dipole correlation factors, presented with the dashed lines in the bottom panel of Fig. 5, suggest that the conformational properties of CB9CB exhibit similar behaviour both as neat compound and after mixing with 5CB. In the isotropic phase $g_{1,iso}^{(intra)}$ is slightly below unity, reflecting weak antiparallel *intra*-dipolar correlations. The onset of the high temperature nematic phase is accompanied by an abrupt enhancement of the more elongated bend conformations, as indicated by the significantly below unity values of $g_{1,\parallel}^{(intra)}$ (green dashed lines in the bottom panel of Fig. 5 and Fig. 7a,b). An interesting feature of the temperature dependence of $g_{1,\parallel}^{(intra)}$ is that after the initial drop at the NI phase transition and the weaker temperature variation within the N phase, it remains practically unaffected by the N-$N_x$ phase transition and saturates at low temperatures to values close to 0.5 both in the neat dimer system, as well as in its mixtures with 5CB. The behaviour of $g_{1,\parallel}^{(intra)}$ in the $N_x$ phase can be explained in terms of a fine interplay between the temperature dependence of the tilt of the phase and bend angle of the dimer, schematically represented in Fig. 7c,d. Specifically, the increase of the tilt angle upon cooling, results in a reduction of the term $\langle (\widehat{L}_1 \cdot \widehat{h})(\widehat{L}_2 \cdot \widehat{h}) \rangle_b$, while the simultaneous increase of the bend angle acts in the opposite manner.

In the high temperature N phase, the intramolecular $g_{1,\perp}^{(intra)}$ factor increases continuously with temperature and, deeply in the N phase, assumes values slightly higher than unity. This is consistent with a continuous increase of the average bend angle on lowering the temperature. The slightly above unity values of $g_{1,\perp}^{(intra)}$ deeply in the N phase drop below unity in the $N_x$ phase, exhibiting a continuous reduction on further cooling. These trends can be interpreted considering the impact of the tilt and bend angle on the relevant term $\langle (\widehat{L}_1 \cdot \widehat{h}_\perp)(\widehat{L}_2 \cdot \widehat{h}_\perp) \rangle_b$, where $\widehat{h}_\perp$ denotes the direction perpendicular to the symmetry axis of the $N_x$ phase. In this case the simultaneous increase of tilt and bend angle on cooling, facilitates the continuous drop of this term, which eventually becomes negative (see Fig. 7).

Summarising our findings on the intramolecular g-factors, we conclude that the tendency of the dimers to adopt more extended conformations continues as the temperature drops. However, the overall increase of the molecular bend-angle on cooling is expressed differently on the projections of the intramolecular g-factors in the direction parallel and perpendicular to the symmetry axis of the nematic phases. In the high temperature N phase, the intramolecular $g_{1,\perp}^{(intra)}$ factor assumes progressively increasing values towards unity, indicating the absence of significant intramolecular dipolar correlations perpendicular to the director deep in the N phase. The onset of the $N_x$ phase has a stronger impact on $g_{1,\perp}^{(intra)}$, than on $g_{1,\parallel}^{(intra)}$, with the former exhibiting a continuous reduction on cooling, assuming eventually values below unity.

*Inter*-molecular dipolar correlations

The successful reproduction of the experimentally measured dielectric permittivities of neat CB9CB and DM mixtures (Fig. 5) is obtained only after the inter-molecular dipole-dipole correlations are considered. From the temperature dependence of $g_{1,\perp(\parallel)}^{(inter)}$, it is evident that the neat CB9CB system exhibits strong antiparallel dipole correlations in both directions, parallel and perpendicular to the symmetry axes of the nematic phases. More specifically, the well below unity values of both intermolecular $g_1$-factors in CB9CB is indicative, as in the case of 5CB, of the antiparallel end-to-end dipolar association of CN units and of the simultaneously synclinic orientational coupling of neighbouring CB units belonging to different particles.

Notably, the fact that $g_{1,\perp(\parallel)}^{(inter)} < 1$ indicates, that side by side nearest-neighbour intermolecular arrangements corresponding to pairs with coaligned steric dipoles (the arrows of the $C_{2v}$ symmetric bend-shaped CB9CB dimers) are less favoured. In the opposite case, both g-factors would assume values greater than unity, as result of the orientational constraints imposed to a pair of bent shaped particles, when they are packed having their steric dipoles parallel. Such orientational correlations, weak in the flexible dimers, are dominant in achiral rigid bent-core mesogens and are connected with smecticity and, possibly, with the stabilisation of polar smectic ordering.[75]

The clearly observed relatively strong dipolar association in the nematic phases is compatible with the formation of chains of dimers having their dipoles in the CB groups antiparallel. The extend of this temperature dependent dynamic physical "polymerisation" becomes limited on increasing 5CB content. Clearly, once the CN unit of a 5CB molecule is involved in the pairing, the "polymerisation" terminates. In the limiting case of the neat 5CB system, this type of physical association is restricted exclusively to pairs of antiparallel associated monomers, and consequently its extend is not particularly sensitive in temperature variations. For this reason, the temperature dependence of g-factors becomes weaker as the concentration in 5CB increases, tending to the temperature independent behaviour of the g-factors of neat 5CB system. This finding does not suggest a different type of local orientational correlations as a function of concentration of 5CB, but rather reflects the extent of the physical polymerisation.

**Relationship of Kirkwood g-factors with measured permittivities**

Clearly, the jump of the dielectric permittivities at the NI phase transition is associated with the onset of the long range orientational order, as well as with an enhancement of the elongated molecular conformations. The variation, however, of the intramolecular correlations with temperature are not sufficient to reproduce the measured permittivities in the nematic phases. For instance, reproduction of the distinctive strong reduction of $\varepsilon_\parallel$ in the nematic phases of CB9CB and its mixtures, requires the consideration of strong antiparallel intermolecular association of neighbouring CN dipoles. On the other hand, the relatively strong variation of $\varepsilon_\perp$ at the N-$N_x$ phase transition is associated mainly with the onset of the tilt of the direction of maximal alignment of the mesogenic units with respect to the axis of modulation and is not connected with a different mode of local dipolar correlations. Concerning the concentration dependence of dielectric permittivity, we conclude that the observed temperature dependence of $\varepsilon_\parallel$ in DM mixtures can be rationalized by the weaker temperature variation of $g_{1,\parallel}^{(inter)}$ on increasing 5CB content, along with the almost concentration independent $g_{1,\parallel}^{(intra)}$ of the dimer.

## Conclusions

We have carried out detailed dielectric and birefringence measurements in the isotropic and the nematic phases of selected mixtures of CB9CB with 5CB. The temperature and composition dependence of the static dielectric permittivity is interpreted in the framework of an anisotropic version of the Kirkwood-Fröhlich theory of dielectric permittivity. The introduction of a molecular field theory, which takes into account the molecular flexibility of the dimers, along with the estimated tilt angle from birefringence measurements, allows for the determination of the temperature and composition dependence of the orientational order parameter of cyanobiphenyl groups, as well as of the conformational statistics of the dimer and the corresponding *intra*-molecular dipolar correlation factors. *Inter*-molecular Kirkwood correlation factors were quantified through fitting of the experimental results to the theoretical model. Antiparallel intermolecular dipole-dipole correlations dominate in both nematic phases of CB9CB and DM mixtures, along and perpendicular to the symmetry axis of each phase. Excellent agreement with the dielectric permittivity measurements, across the whole temperature and composition ranges, manifests the necessity of both *intra*- and *inter*-molecular dipolar associations for the quantitative description of the dielectric permittivity of dimeric liquid crystals and provides valuable insight into the supramolecular organisation within the nematic phases of these systems.

## Acknowledgements

The authors acknowledge Prof. C. A. Krontiras (University of Patras, Greece) for assistance with magnetic field measurements and Dr. Ioannis Raptis (Research Director of INN-NCSR Demokritos, Theta Metrisis) and Mrs. Savvina Fournari (Theta Metrisis) for WLRS measurements. E. Z. acknowledges financial support by the Hellenic Foundation for Research and Innovation (HFRI PhD Fellowship grant, Fellowship Number: 80996). C. W. and Z. A. acknowledge support by the EPSRC (UK) through the project EP/M015726/1.

Evangelia E. Zavvou *[1], Efthymia Ramou [1], Ziauddin Ahmed [2], Chris Welch [2], Georg H. Mehl [2], Alexandros G. Vanakaras [3] and Panagiota K. Karahaliou [1]

[1] Department of Physics, University of Patras, 26504 Patras, Greece.
[2] Department of Chemistry, University of Hull, HU6 7RX, UK.
[3] Department of Materials Science, University of Patras, 26504 Patras, Greece.

I. POM textures and temperature dependence of the dielectric permittivity

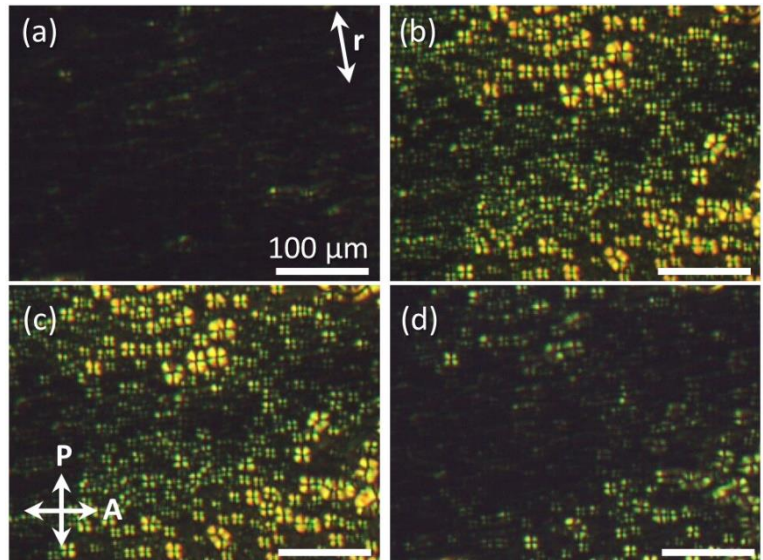

*Figure SI1.* Optical textures of the $N_x$ phase obtained during POM observations following Protocol 2 (see main text); Cooling in the presence of 25 $V_{rms}$ at (a) 70.5°C and (b) 48°C and subsequent heating after field removal at (c) 58.3°C and (d) 70.5°C.

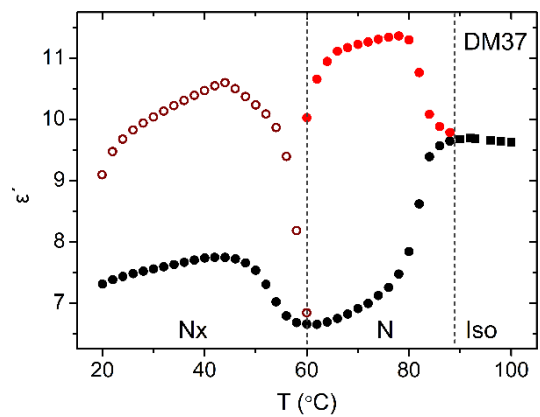

*Figure SI2.* Temperature dependence of dielectric permittivity at $f$ = 5 kHz of DM37. Squares-$\varepsilon_{iso}$; Black circles-$\varepsilon_\perp$; Red circles-$\varepsilon_\parallel$ measured with P.1; Open circles–$\varepsilon_\parallel$ measured with P.2. The progressive loss of homeotropic alignment achieved with Protocol 2 is observed on heating towards the N phase.

## II. Conformational statistics of the dimer molecules

The conformational properties of a single isolated symmetric dimer are associated with the potential $E(n; \hat{L}_1, \hat{L}_2)$ which takes into account both bonded and non-bonded intramolecular interactions. Here $n$ defines the molecular conformation, or, equivalently, describes a conformational state through the necessary internal degrees of freedom of the linear flexible spacer, i.e. torsion and bond angles, bond lengths etc., and $\hat{L}_1, \hat{L}_2$ correspond to the conformation dependent orientations of the rigid mesogenic cores which are assumed cylindrically symmetric, see inset in Figure SI3.

The temperature dependent probability density, $p(c) = \langle \delta(c - \hat{L}_1 \cdot \hat{L}_2) \rangle_{conf}$, i.e. the probability to find the rigid cores of a single isolated dimer forming an angle $\theta_{12}$ with $cos\theta_{12} = c$, is of central importance for the study of orientational dependent properties in the nematic state. In this work we have modelled, $p(c)$, as $p(c) = \zeta \exp[-\beta E(c)]$ where $\beta = 1/k_B T$, $\zeta$ the normalization constant and $E(c)$ an effective internal energy of the dimer as a function of the relative orientation of the cores. Formally, $E(c)$ can be represented as a Legendre polynomial series $E(c) = \sum_{i=0}^{\infty} a_l P_l(c)$ with a clear dependence of the expansion coefficients on the details of the intramolecular potential. The first three terms of the expansion are enough to capture the essential features of the $p(c)$ that have already been calculated for symmetric odd and/or even dimers at various degrees of resolution concerning the generation of the dominant molecular conformations of the spacer.[1–4] The dominance of the $P_1 - P_3$ terms on the behaviour of the orientational coupling of the bonded mesogens is an intrinsic property of these dimers and not the circumstantial outcome of a particular modelling.

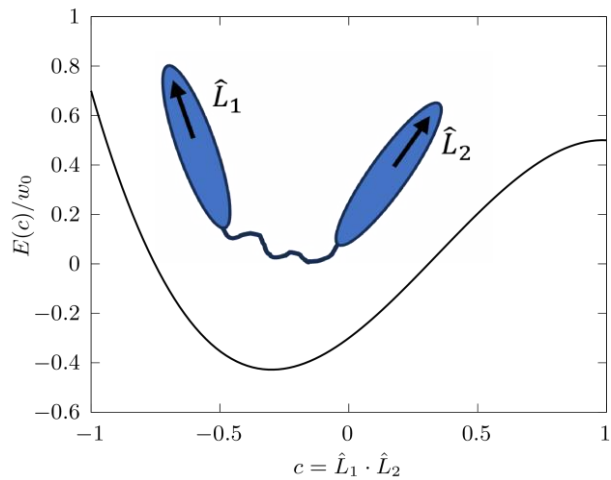

**Figure SI3.** *Internal energy of the dimer as a function of the relative orientation of the cores. A cartoon of the dimer is presented indicating the orientation of the molecular dipoles in the mesogenic cores.*

Using as the only inputs the statistics of the conformation of the dimer and the experimental values of $T_{NI}$, the Maier-Saupe mean field theory described in the manuscript provides the self-consistency relationships for the calculation of the full orientational/conformational distribution function of the dimer (and of its mixtures with its monomer), as well as the free energy of the systems at constant volume/composition conditions.

In our calculations for CB9CB we have chosen the set of the $a_l$ parameters as $a_1 = 0.25w_0$, $a_2 = 0.6w_0$ and $a_3 = -0.35w_0$, with $w_0/k_B \approx 1398K$, see the main text of the manuscript for details. With this parameterisation the effective intramolecular potential $E(c)$ is presented in Figure SI3.

The statistics of conformations of the flexible dimers change in the nematic phase according to Eq 10 of the manuscript due to the orientational coupling between the mesogenic units and the nematic director field. The integral of the orientational/conformational dependent probability $f^{(dimer)}(\omega, c)$ over the molecular orientations with respect to the nematic director gives the probability distribution of the angles between the mesogenic units in the bulk, $p_{bulk}^{(dimer)}(c)$. According to Figure SI4, $p_{bulk}^{(dimer)}(c)$ exhibits pronounced changes at the N-I transition which are associated with the onset of long range orientational order. Comparing the distribution in the isotropic phase (red dashed curve) with the corresponding distribution with the onset of the nematic phase (thick black curve), it becomes immediately evident that the maximum probability changes towards molecular states with larger opening angle between the mesogenic units. A noticeable increase of the U-shaped conformations is also observed. These trends continue in the nematic phase as the temperature decreases (thin grey lines). These are exactly the same trends observed in other works as well, see for instance Fig. 3 in ref. 1, where the corresponding probability distribution is referred to a dimer with 11 carbons in the spacer.

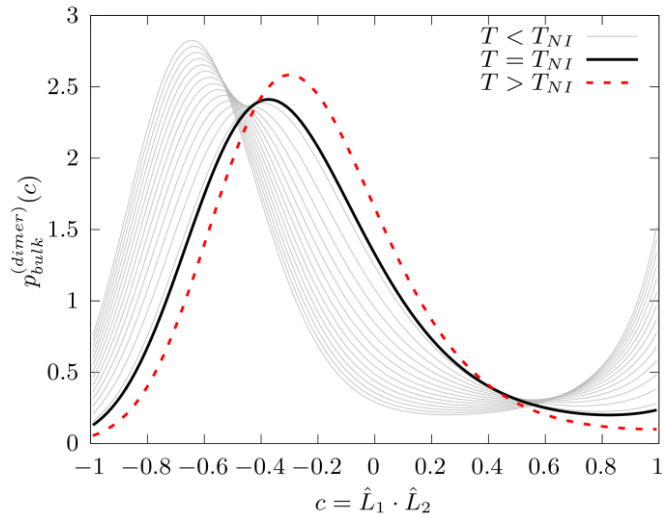

*Figure SI4. Probability distribution $p_{bulk}^{(dimer)}(c)$ of the model CB9CB dimer in the bulk, calculated at several temperatures. The dashed red line corresponds to a temperature slightly above $T_{NI}$. The thick black line gives the distribution at the onset of the nematic state, while the thin grey lines are the corresponding distributions calculated lowering the temperature at steps of $5K$.*

The net molecular dipole moment due to the permanent dipoles, $\boldsymbol{m}$, of a given molecular conformation having the cyanobiphenyl units pointing in the $\hat{\boldsymbol{L}}_1$ and $\hat{\boldsymbol{L}}_2$ directions, is given by $\boldsymbol{m} = \mu_{CN}(\hat{\boldsymbol{L}}_1 + \hat{\boldsymbol{L}}_2)$ with, $\mu_{CN}$, the electric dipole moment of the CN terminal group. The components of the mean-square molecular dipole along and perpendicular to the nematic director are given by: $m_\parallel^2 \equiv \langle(\boldsymbol{m} \cdot \hat{\boldsymbol{n}})^2\rangle$ and $2m_\perp^2 = \langle m^2 \rangle - m_\parallel^2$, with $\langle m^2 \rangle = 2(1 + \langle \hat{\boldsymbol{L}}_1 \cdot \hat{\boldsymbol{L}}_2 \rangle)$.

We have $\langle(\mathbf{m}\cdot\hat{\mathbf{n}})^2\rangle = \mu_{CN}^2\left(\langle(\hat{\mathbf{L}}_1\cdot\hat{\mathbf{n}})^2 + (\hat{\mathbf{L}}_2\cdot\hat{\mathbf{n}})^2\rangle + 2(\hat{\mathbf{L}}_1\cdot\hat{\mathbf{n}})(\hat{\mathbf{L}}_2\cdot\hat{\mathbf{n}})\right)$, which can be rewritten as $\langle(\mathbf{m}\cdot\hat{\mathbf{n}})^2\rangle = \mu_{CN}^2\langle(\hat{\mathbf{L}}_1\cdot\hat{\mathbf{n}})^2 + (\hat{\mathbf{L}}_2\cdot\hat{\mathbf{n}})^2\rangle\left(1 + 2\frac{\langle(\hat{\mathbf{L}}_1\cdot\hat{\mathbf{n}})(\hat{\mathbf{L}}_2\cdot\hat{\mathbf{n}})\rangle}{\langle(\hat{\mathbf{L}}_1\cdot\hat{\mathbf{n}})^2 + (\hat{\mathbf{L}}_2\cdot\hat{\mathbf{n}})^2\rangle}\right)$. Inserting into the last equation the definition of the nematic orientational order parameter associated with the ordering of the CB units of the dimer, $S_d = \langle P_2(\hat{\mathbf{L}}_1\cdot\hat{\mathbf{n}}) + P_2(\hat{\mathbf{L}}_2\cdot\hat{\mathbf{n}})\rangle/2$, we obtain the mean-square of the total molecular dipole along the nematic director, $m_\parallel^2 = 2\mu_{CN}^2\frac{1}{3}(1+2S_d)g_{1,\parallel}^{(intra)}$, $g_{1,\parallel}^{(intra)} = 1 + 3\frac{\langle(\hat{\mathbf{L}}_1\cdot\hat{\mathbf{n}})(\hat{\mathbf{L}}_2\cdot\hat{\mathbf{n}})\rangle}{1+2S_d}$, which are the first two equations in Eq. (7) of the manuscript.

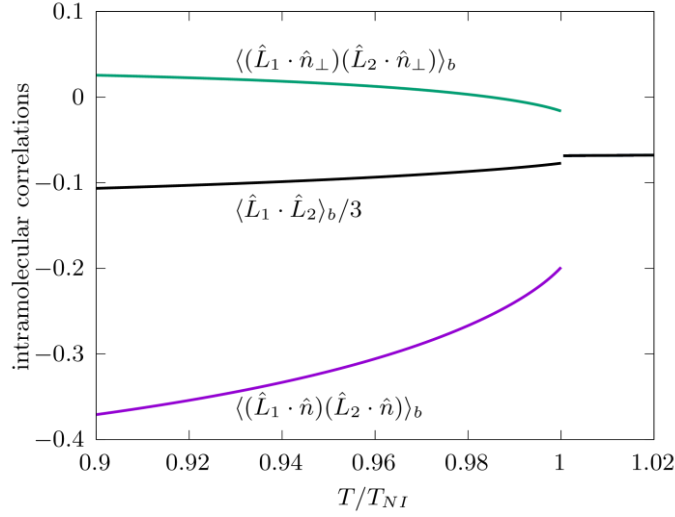

*Figure SI5. Intramolecular orientational correlation factors vs reduced temperature for the neat CB9CB system.*

In Figure SI5 we present the temperature dependence of the various *intra*molecular correlation factors: $\langle\hat{\mathbf{L}}_1\cdot\hat{\mathbf{L}}_2\rangle_b$, $\langle(\hat{\mathbf{L}}_1\cdot\hat{\mathbf{n}})(\hat{\mathbf{L}}_2\cdot\hat{\mathbf{n}})\rangle_b$ and $\langle(\hat{\mathbf{L}}_1\cdot\hat{\mathbf{n}}_\perp)(\hat{\mathbf{L}}_2\cdot\hat{\mathbf{n}}_\perp)\rangle_b$ with $\hat{\mathbf{n}}_\perp$ a direction perpendicular to the nematic director. We note here that the close to zero values of $\langle(\hat{\mathbf{L}}_1\cdot\hat{\mathbf{n}}_\perp)(\hat{\mathbf{L}}_2\cdot\hat{\mathbf{n}}_\perp)\rangle_b$ should not be interpreted as a result of the lack of orientational correlations between the mesogenic cores, but rather as a specific property related to the form of the probability distribution of the relative orientations of the mesogenic cores in the bulk, $p_{bulk}^{(dimer)}(c)$. With the adopted parameterisation, we get for a single CB9CB dimer in an isotropic environment at $T = 400K$, $\langle\hat{\mathbf{L}}_1\cdot\hat{\mathbf{L}}_2\rangle_b \approx -0.2$ indicating a weak antiparallel intramolecular dipolar correlation, note that in the graph we present the temperature dependence of $\langle\hat{\mathbf{L}}_1\cdot\hat{\mathbf{L}}_2\rangle_b/3$. In the nematic phase and as the temperature drops the antiparallel correlation is strongly enhanced as result of the orientational coupling of the mesogenic units with the nematic field. This is clearly demonstrated by the jump of $\langle(\hat{\mathbf{L}}_1\cdot\hat{\mathbf{n}})(\hat{\mathbf{L}}_2\cdot\hat{\mathbf{n}})\rangle_b$ at the NI phase transition and its temperature dependence in the nematic phase (magenta curve in Fig. SI5).

In the graphs of Fig. 4 in the manuscript and in the calculations that will follow, we consider only the high temperature nematic phase which, according to experiment, is stable up to $T \approx 0.95T_{NI}$ for the neat CB9CB system. Bellow this temperature the nano-modulated nematic, $N_x$, becomes stable. See the main text of the manuscript on how the experimentally calculated temperature dependence of birefringence can be utilised to allow the calculation of the relevant order parameters and orientational correlations in the low temperature nematic phase. In Figure SI6 we present the temperature dependence of (a) the orientational order parameter $S_d$ of the dimer which, by definition, is the order parameter associated with the ordering of the CB groups of the molecule, and (b) the mean square components of the scaled dipole moment, $m_\lambda^2/\mu_{CN}^2$. From Figure SI6(b) it is evident, that while $m_\perp^2$ decreases with decreasing temperature in the nematic phase, the parallel component $m_\parallel^2$, remains relative insensitive to temperature variations. In the isotropic phase $m^2$ remains practically constant. While the temperature dependence of $m_\perp^2$ is in qualitative agreement with the behaviour of $\varepsilon_\perp$ (dielectric constant perpendicular to the nematic director), the temperature dependence of $m_\parallel^2$ is in remarkable disagreement with the experimentally measured $\varepsilon_\parallel(T)$, which after a small jump at the NI phase transition drops rapidly upon lowering the temperature. In addition, the practically constant mean-square dipole moment in the isotropic phase is also in disagreement with the relative strong variation of the measured dielectric constant with temperature in the isotropic phase, see Fig. 4(a) in the manuscript. These observations, suggest that the ignored, up to this point, *inter*molecular dipole-dipole correlations, i.e. considering $g_{1,\parallel(\perp)}^{(inter)} = 1$ in Eq. (5) of the manuscript, should be taken into consideration for the calculation of the components $[\mu^2]_{\parallel(\perp)} = m_{\parallel(\perp)}^2 g_{1,\parallel(\perp)}^{(inter)}$ of the actual mean square dipole moment. Actually, this is one of the main topics of the manuscript.

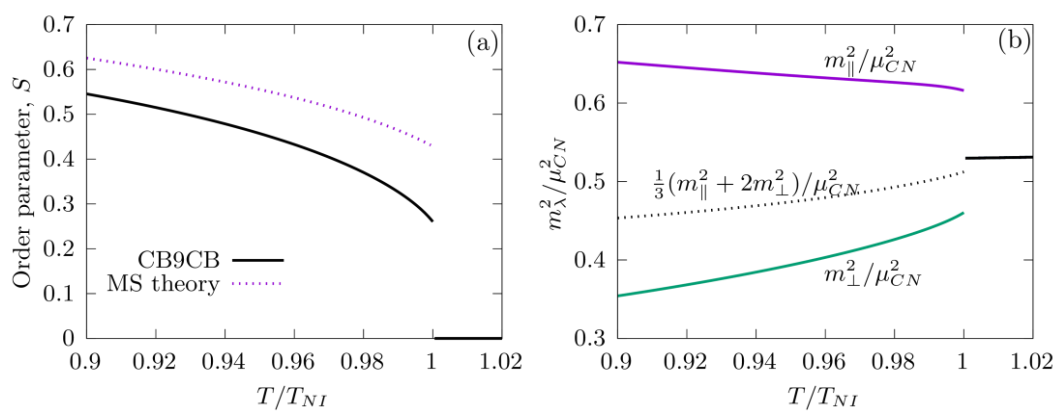

*Figure SI6.* (a) Orientational order parameter of the neat CB9CB dimer system vs reduced temperature (solid line). For comparison in the same plot the universal Maier-Saupe nematic order parameter for rigid rod-like mesogens is also presented (dotted line). (b) Mean square components of the scaled dipole moment vs reduced temperature (solid lines); the dotted line is the average mean square dipole moment.